\documentstyle[12pt]{article}
\begin{document}

\begin{titlepage}
\begin{center}
\hspace*{10cm} Preprint IFUNAM\\
\hspace*{10cm} FT-93-021\\
\hspace*{10cm} June 1993\\
\vspace*{35mm}

{\Large{\bf ELECTRODYNAMICS\\
\vskip1mm
WITH WEINBERG'S PHOTONS}}\\
\vspace*{10mm}
{\large{\bf Valeri V. Dvoeglazov}}$^{*,\,\dagger}$\\
\vskip1mm
{\it Depto de F\'{\i}s.Te\'orica, Instituto de F\'{\i}sica\\
   Universidad Nacional Aut\'onoma de Mexico\\
   Apartado Postal 20-364, 01000 Mexico, D.F. MEXICO}\\
\vspace*{3mm}
\end{center}
\vspace*{2mm}
\begin{abstract}

The interaction of the spinor field with the Weinberg's $2(2S+1)$- component
massless field is considered. New interpretation of the
Weinberg's spinor is proposed. The equation analogous to the Dirac oscillator
is obtained.
\end{abstract}

\vspace*{25mm}
\vspace*{25mm}


\vspace*{-2mm}
\noindent
------------------------------------------------------------------------------------------------\\
\footnotesize{
\noindent
$^{*}$ On leave from: {\it Dept. Theor.} \& {\it Nucl. Phys.,
Saratov State University and Sci.} \& {\it Tech. Center for  Control
and Use of Physical Fields and Radiations, Astrakhanskaya str. , 83,\,\,
Saratov 410071 RUSSIA}\\
\noindent
$^{\dagger}$ Email: valeri@ifunam.ifisicacu.unam.mx\\
\hspace*{15mm}dvoeglazov@main1.jinr.dubna.su}\\
\end{titlepage}

\newpage
\normalsize{
In the Weinberg's $2(2S+1)$- approach~\cite{Weinberg} to  description of
particles of spin $S=1$ the wave function (WF)  of vector bosons is written as
six component column. It satisfies the following motion equation:
\begin{equation}
\left[\gamma_{\mu\nu} p_{\mu} p_{\nu}+M^2\right] \Psi^{(S=1)} (x) = 0,
\end{equation}
where $\gamma_{\mu\nu}$ 's are  covariantly defined $6\otimes6$-
matrices~\cite{BM}, $\mu,\nu=1\ldots 4$.

The 6- component WF's  $\Psi^{(S=1)}={\Phi \choose \Xi}$ are transformed on the
$(1,0)\oplus(0,1) $ representation of the Lorentz universal covering group
$SL(2,C)$ .  This way of description of the $S=1$ particle has some advantages,
indeed~\cite{Tucker}.

In Ref. [1a, p.B1323] and Ref.~\cite[p.361]{Marinov} the following
invariant (the interaction Hamiltonian)  for interaction of 3 bispinors (e.g.
two particles of spin $S=1/2$ and one particle of spin $S=1$) has been
constructed:
\begin{equation}\label{eq:hm}
{\cal H}_{\Psi\psi\psi}=g\sum_{\mu_1\,\mu_2\,\mu_3}\left (\matrix{
S_1 & S_2 & S_3 \cr
\mu_1 & \mu_2 & \mu_3 \cr
}\right ) \Phi^{\mu_1}_{(S_1)}\phi^{\mu_2}_{(S_2)}\phi^{\mu_3}_{(S_3)}
\pm \left (\matrix{
S_1 & S_2 & S_3 \cr
\dot{\mu}_1 &\dot{\mu}_2 & \dot{\mu}_3 \cr
}\right )
\Xi^{\dot{\mu}_1}_{(S_1)}\chi^{\dot{\mu}_2}_{(S_2)}\chi^{\dot{\mu}_3}_{(S_3)},
\end{equation}
\\
where
$\left (\matrix{
S_1 & S_2 & S_3 \cr
\mu_1 & \mu_2 & \mu_3 \cr
}\right )$
are the Wigner $3j$- symbols.\\

Assuming the interpretation of the Weinberg's spinor as the sum of vector and
pseudovector [5b, formula (5)]\footnote{ In Ref. ~\cite{Cabbibo} to  the
importance of the pseudovector potential $\tilde{A}_k$ in QED has been paid
attention.}$^{,}\,$\footnote{As shown in Ref. [5a] the interpretation
$\Psi^{(S=1)}$ according to [1b, p.B888] leads to the contradiction with the
theorem about connection between the $(A,B)$ representation of the Lorentz
group and the helicity of particle with the WF which transforms according to
this representation ($B-A=\lambda$). Moreover, the Weinberg's equations
[1b, formulas (4.21) and (4.22)] admit the acausal ($E\neq\pm p$)
solutions~\cite{Ahlu}.}:
\begin{eqnarray}
\cases{\Phi_{k} = \tilde{A}_{k} + iA_{k},& $ $\cr
\Xi_{k} = \tilde{A}_{k} - iA_{k}.& $ $}
\end{eqnarray}
in the case of massless  $S=1$ particles (photons) we get the following
invariant for interaction of two spinor particles with the electromagnetic
field (the spinor representation is used) :
\begin{eqnarray}\label{eq:lagr}
\lefteqn{{\cal H}_{\Psi\psi\psi}=g\sum_{k\,\mu_2\,\mu_3}\left\{\left [\left
(\matrix{
1 & {1\over 2} & {1\over 2} \cr
k & \mu_2 & \mu_3 \cr
}\right ) \phi^{\mu_2}_{({1\over 2})}\phi^{\mu_3}_{({1\over 2})}
+ \left (\matrix{
1 & {1\over 2} & {1\over 2} \cr
k &\dot{\mu}_2 & \dot{\mu}_3 \cr
}\right )
\chi^{\dot{\mu}_2}_{({1\over 2})}\chi^{\dot{\mu}_3}_{({1\over 2})}\right
]\tilde{A}_k+\right.}\nonumber\\
&+&i\left.\left [ \left (\matrix{
1 & {1\over 2} & {1\over 2} \cr
k & \mu_2 & \mu_3 \cr
}\right )\phi^{\mu_2}_{({1\over 2})}\phi^{\mu_3}_{({1\over 2})}
- \left (\matrix{
1 & {1\over 2} & {1\over 2} \cr
k &\dot{\mu}_2 & \dot{\mu}_3 \cr
}\right )
\chi^{\dot{\mu}_2}_{({1\over 2})}\chi^{\dot{\mu}_3}_{({1\over 2})}\right
]A_k\right\}.
\end{eqnarray}
( In (\ref{eq:hm}) we choose the sign $"+"$ for definity).
Taken into account the relation between the Pauli $\sigma$-- matrices and the
Clebsh-Gordon coefficients (formula on the p. 65 in
\begin{equation}
\sigma^{\mu}_{\alpha\beta}=-\sqrt{3} C^{{1\over 2}\alpha}_{1\mu{1\over 2}\beta}
\end{equation}
one can rewrite the previous expression (\ref{eq:lagr}) as following:
\begin{equation}\label{eq:Ham}
{\cal H}_{\Psi\bar\psi\psi}= \frac{g}{\sqrt{6}}\left \{-\bar
\psi\alpha_k\gamma_5 \psi\tilde{A}_k+i\bar\psi\alpha_k \psi A_k\right \}.
\end{equation}
In fact, the coupling constant $g$ is equal to $ie\sqrt{6}$, $e$ is electric
charge,\, $k=1,2,3$. The matrix  $\gamma_5$ has been chosen in the diagonal
form:
\begin{equation}
\gamma_5=\pmatrix{
-1 & 0 \cr
0 & 1 \cr
},
\end{equation}
\begin{equation}
\beta=\alpha_4=\pmatrix{
0 & 1 \cr
1 & 0 \cr
},
\end{equation}
and
\begin{equation}
\vec\alpha=\pmatrix{
\vec\sigma & 0 \cr
0 & -\vec\sigma \cr
}.
\end{equation}

One can see that this interaction Hamiltonian leads to the equations
(\ref{eq:osc}), which are analogous to the Dirac oscillator
equations~\cite{Moshinsky} provided that we suppose
$\tilde{A}_k=m\omega r_k/e$.
In fact, we have the Eq. from the Hamiltonian (\ref{eq:Ham}):
\begin{equation}\label{eq:Dir}
i\hbar\frac{\partial\psi}{\partial t}=c\vec\alpha\cdot(\vec p-e\vec
A-ie\gamma_5\vec{\tilde{A}})\psi+mc^2\beta\psi,
\end{equation}
which is equivalent to the following system ($c=\hbar=1$) :
\begin{eqnarray}
\cases{\left [(\vec\sigma\vec p)-e(\vec \sigma\vec
A)+ie(\vec\sigma\vec{\tilde{A}})\right ]\xi+m\eta=E\xi,& $ $\cr
\left [-(\vec\sigma\vec p)+e(\vec\sigma\vec
A)+ie(\vec\sigma\vec{\tilde{A}})\right ]\eta+m\xi=E\eta.& $ $}
\end{eqnarray}
Therefore,
\begin{eqnarray}
\lefteqn{(E^2-m^2)\xi=\left \{\vec p\,^2-e\left [(\vec \sigma\vec p)(\vec
\sigma\vec A)+(\vec \sigma\vec A)(\vec\sigma\vec p)\right ]+\right.}\nonumber\\
&+&\left. ie\left [(\vec\sigma\vec
p)(\vec\sigma\vec{\tilde{A}})-(\vec\sigma\vec{\tilde{A}})(\vec\sigma\vec
p)\right ]+e^2\vec
A\,^2+e^2\vec{\tilde{A}}\,^2+2ieE(\vec\sigma\vec{\tilde{A}})\right \}\xi,
\end{eqnarray}
and
\begin{eqnarray}
\lefteqn{(E^2-m^2)\eta=\left \{\vec p\,^2-e\left [(\vec \sigma\vec p)(\vec
\sigma\vec A)+(\vec \sigma\vec A)(\vec\sigma\vec p)\right ]-\right.}\nonumber\\
&-&\left. ie\left [(\vec\sigma\vec
p)(\vec\sigma\vec{\tilde{A}})-(\vec\sigma\vec{\tilde{A}})(\vec\sigma\vec
p)\right ]+e^2\vec
A\,^2+e^2\vec{\tilde{A}}\,^2+2ieE(\vec\sigma\vec{\tilde{A}})\right \}\eta,
\end{eqnarray}
what give (when $\vec A=\vec 0$ and $\vec{\tilde{A}}=m\omega \vec r/e$):
\begin{eqnarray}\label{eq:osc}
\cases{(E^2-m^2)\xi=\left [\vec p\,^2+m^2\omega^2 r^2+2iEm\omega(\vec\sigma\vec
r)+3m\omega+4m\omega \vec S[\vec r \times \vec p]\right ]\xi,& $ $\cr
(E^2-m^2)\eta=\left [\vec p\,^2+m^2\omega^2 r^2+2iEm\omega(\vec\sigma\vec
r)-3m\omega-4m\omega \vec S[\vec r \times \vec p]\right ]\eta.& $ $}
\end{eqnarray}

The appearance of new term ($2iEm\omega(\vec\sigma\vec r)$)  can be explained
by the fact that it is possible to add in the formula (5) of the
paper~\cite{Moshinsky} both the term $-im\omega \beta\vec r$, which corresponds
to the addition $\alpha_i\wedge\alpha_4 R_{4i}$ (where $R_{4j}=ir_j$), and the
one ${m\omega\over 2} [\vec\alpha\times~\vec r]$, which corresponds to the
interaction term ${1\over 2}
\alpha_i\wedge\alpha_j R_{ij}$ (where $R_{ij}=\epsilon_{ijk}r_k$), in
accordance with bivector construction rules as the expansion  in Clifford
algebra in the Minkowsky 4- dimensional space~\cite{Jancewicz}.
So, the interaction term for the Dirac oscillator is possible to define:
\begin{equation}
R=-m\omega\alpha_i\wedge\alpha_4 R_{4i}+\frac{m\omega}{2}\alpha_i\wedge\alpha_j
R_{ij}.
\end{equation}
(cf. formula (32) in Ref.~\cite[p. 244]{Jancewicz}). In the case of
electromagnetism $R_{4i}=i\tilde{A}_i$ and $R_{ij}=\epsilon_{ijk}\tilde{A}_k$.
Thus, instead of the minimal form of electromagnetic interaction ($\gamma_\mu
A_\mu$) we have the bivector form interaction (similarly to the introduction of
the Pauli term).

In this case the Eqs. (7a,7b) in Ref.~\cite{Moshinsky} could be written in the
form (standard representation)\footnote{Since the Eq. (\ref{eq:Dir}) does not
lead to the term \, $-im\omega\beta\vec r$ this induces us to the assumption of
existence of the "irregular" invariant for the interaction of two $S=1/2$
particles and one $S=1$ particle (cf. formula (15) in Ref.~\cite{Marinov}).}:
\begin{eqnarray}
(E-mc^2)\psi_1&=&c\vec\sigma\cdot(\vec p+im\omega\vec
r)\psi_2+icm\omega(\vec\sigma\vec r)\psi_1,\nonumber\\
(E+mc^2)\psi_2&=&c\vec\sigma\cdot(\vec p-im\omega\vec
r)\psi_1+icm\omega(\vec\sigma\vec r)\psi_2.
\end{eqnarray}

Would like to mention that $A_k$, the vector potential, is the compensating
field for the gauge transformation of the second kind and $\tilde{A}_k$ , the
pseudovector potential, is the compensating field for the chirality
transformation\footnote{See e.g.~\cite{Strazhev} for discussion of the
chirality ($\gamma_5$) symmetry of massless fields and neutrino theory of
photons.  As to the generalized gauge transformations, one can find
in~\cite{Barut, Crawford}.}.

Since  $E_k=rot \tilde{A}_k$ we can see that $\vec{E}=\vec{0}$, and
$\vec{H}=\vec{0}$ in this case. However, the spectrum is influenced by the term
$\vec{\tilde{A}}$. Perhaps this situation is linked somehow with the
Aharonov-Bohm effect.

We can implement the new $4\otimes 4$- matrix field corresponding to the
electromagnetic field:
\begin{equation}
\Phi_k=\pmatrix{
A_k-i\tilde{A}_k & 0 \cr
0 & A_k+i\tilde{A}_k  \cr
}
\end{equation}
which is described by the Lagrangian:
\begin{equation}
{\cal L}=\bar\Psi^{(S=1)}\gamma_{\mu\nu}p_{\mu}p_{\nu}\Psi^{(S=1)}=
i\bar\Phi_j\left \{-i\epsilon_{ijk}p_4 p_i\otimes \gamma_5+(\vec
p\,^2\delta_{jk}-p_j p_k)\otimes I\right \}\Phi_k.
\end{equation}
The corresponding dynamical invariants are found from the tensor of
energy-momentum which is written as following:
\begin{eqnarray}
T_{44}&=&i\bar\Phi_j(\vec p\,^2\delta_{jk}-p_j p_k) \Phi_k,\nonumber\\
T_{\,l4}&=&i\epsilon_{ijk}\bar\Phi_j p_i p_l\otimes\gamma_5\,\Phi_k,\nonumber\\
T_{\,4l}&=&i\epsilon_{ljk}\bar\Phi_j p_4 p_4\otimes \gamma_5
\,\Phi_k-2i\bar\Phi_k p_l p_4\Phi_k+i\bar\Phi_k p_k p_4 \Phi_l+i\bar\Phi_l p_4
p_k\Phi_k,\nonumber\\
T_{\,lm}&=&{\cal L}\delta_{lm}+i\epsilon_{mjk}\bar\Phi_j p_l
p_4\otimes\gamma_5\,\Phi_k - 2i\bar \Phi_k p_l p_m\Phi_k+i\bar\Phi_m p_l
p_k\Phi_k+i\bar\Phi_k p_k p_l\Phi_m.
\end{eqnarray}

The problem of quantization of this field will be considered in the approaching
publication.

The author expresses his gratitude to Profs.  D. Ahluwalia, R. N. Faustov, V.
G. Kadyshevsky,  M. Moshinsky and Yu. F. Smirnov
for extremely fruitful discussions. The technical assistance of A. S. Rodin is
greatly appreciated. The author also appreciates very much excellent working
conditions when staying at the Instituto de F\'{\i}sica, UNAM. This work has
been financially supported by the CONACYT (Mexico) under the contract No.
920193.
}

\end{document}